
\footline={\tenrm\hss\folio\hss}

\def\date{August, 1985; revised September, 1987, and May, 1988}

\font\ninerm=cmr9   \font\eightrm=cmr8   \font\sixrm=cmr6
\font\ninei=cmmi9      \font\sixi=cmmi6
\font\ninesy=cmsy9    \font\sixsy=cmsy6
\font\ninebf=cmbx9    \font\sixbf=cmbx6
\font\ninett=cmtt9  
\font\nineit=cmti9  
\font\ninesl=cmsl9  

\font\titless=cmbx10 scaled \magstep2

\catcode`@=11
\newskip\ttglue

\def\tenpoint{\def\rm{\fam0\tenrm}
  \textfont0=\tenrm \scriptfont0=\sevenrm \scriptscriptfont0=\fiverm
  \textfont1=\teni  \scriptfont1=\seveni  \scriptscriptfont1=\fivei
  \textfont2=\tensy \scriptfont2=\sevensy \scriptscriptfont2=\fivesy
  \textfont3=\tenex \scriptfont3=\tenex   \scriptscriptfont3=\tenex
  \textfont\itfam=\tenit  \def\it{\fam\itfam\tenit}%
  \textfont\slfam=\tensl  \def\sl{\fam\slfam\tensl}%
  \textfont\ttfam=\tentt  \def\tt{\fam\ttfam\tentt}%
  \textfont\bffam=\tenbf  \scriptfont\bffam=\sevenbf
   \scriptscriptfont\bffam=\fivebf  \def\bf{\fam\bffam\tenbf}%
  \tt \ttglue=.5em plus .25em minus.15em
  \parindent=20pt
  \smallskipamount=3pt plus1pt minus1pt
  \medskipamount=6pt plus2pt minus2pt
  \bigskipamount=12pt plus4pt minus4pt
  \normalbaselineskip=12pt
  \setbox\strutbox=\hbox{\vrule height8.5pt depth3.5pt width0pt}%
  \let\sc=\eightrm  \let\big=\tenbig  \normalbaselines\rm}

\def\ninepoint{\def\rm{\fam0\ninerm}
  \textfont0=\ninerm \scriptfont0=\sixrm \scriptscriptfont0=\fiverm
  \textfont1=\ninei  \scriptfont1=\sixi  \scriptscriptfont1=\fivei
  \textfont2=\ninesy \scriptfont2=\sixsy \scriptscriptfont2=\fivesy
  \textfont3=\tenex \scriptfont3=\tenex   \scriptscriptfont3=\tenex
  \textfont\itfam=\nineit  \def\it{\fam\itfam\nineit}%
  \textfont\slfam=\ninesl  \def\sl{\fam\slfam\ninesl}%
  \textfont\ttfam=\ninett  \def\tt{\fam\ttfam\ninett}%
  \textfont\bffam=\ninebf  \scriptfont\bffam=\sixbf
   \scriptscriptfont\bffam=\fivebf  \def\bf{\fam\bffam\ninebf}%
  \tt \ttglue=.5em plus .25em minus.15em
  \parindent=20pt
  \smallskipamount=3pt plus1pt minus1pt
  \medskipamount=6pt plus2pt minus2pt
  \bigskipamount=12pt plus4pt minus4pt
  \normalbaselineskip=11pt
  \setbox\strutbox=\hbox{\vrule height8pt depth3pt width0pt}%
  \let\sc=\sevenrm  \let\big=\ninebig  \normalbaselines\rm}

\def\tenbig#1{{\hbox{$\left#1\vbox to8.5pt{}\right.\n@space$}}}
\def\ninebig#1{{\hbox{$\textfont0=\tenrm\textfont2=\tensy
  \left#1\vbox to7.25pt{}\right.\n@space$}}}
\def\eightbig#1{{\hbox{$\textfont0=\ninerm\textfont2=\ninesy
  \left#1\vbox to6.5pt{}\right.\n@space$}}}

\def\famzero{\fam\z@}

\def\tit{\it}
\def\emph{\tit}  \def\demph{\tit} 

\def\bb{3pc}
\def\cc{2pc}
\def\dd{.5pc}

\def\sectionskip{\penalty-200\vskip \bb plus 12pt minus 6pt}
\def\secbegin#1//{\centerline{\bf#1}\penalty1000\vskip \cc plus 6pt minus 4pt\penalty1000}
\def\pfbegin#1.{\penalty25\noindent{\tit#1:}\xskip\ignorespaces}
\def\xskip{\hskip 7pt plus 3pt minus 4pt}

\def\QEDh{\hbox{\vrule height 6pt\ignorespaces
                \vbox to 6pt{\hrule width 5.2pt
                             \vfill
                             \hrule width 5.2pt}\ignorespaces
                \vrule height 6pt}}
\def\qed{{\penalty1000}\quad{\penalty1000}\QEDh} 

\outer\def\jsproclaim #1. #2\par{\medbreak
  \noindent{\bf#1.\enspace}{\it#2}\par
  \ifdim\lastskip<\medskipamount \removelastskip\penalty55\medskip\fi}
\outer\def\beginjsproclaim #1. {\medbreak
  \noindent{\bf#1.\enspace}\begingroup\it}
\outer\def\endjsproclaim{\endgroup\par
  \ifdim\lastskip<\medskipamount \removelastskip\penalty55\medskip\fi}

\outer\def\remark #1. #2\par{\medbreak
  \noindent{\tit#1.\enspace}{#2}\par
  \ifdim\lastskip<\medskipamount \removelastskip\penalty55\medskip\fi}
\outer\def\beginremark #1. {\medbreak
  \noindent{\tit#1.\enspace}\begingroup}
\outer\def\endremark{\endgroup\par
  \ifdim\lastskip<\medskipamount \removelastskip\penalty55\medskip\fi}

\outer\def\defn #1. #2\par{\medbreak
  \noindent{\bf#1.\enspace}{#2}\par
  \ifdim\lastskip<\medskipamount \removelastskip\penalty55\medskip\fi}

\def\today{\ifcase\month\or
  January\or February\or March\or April\or May\or June\or
  July\or August\or September\or October\or November\or December\fi
  \space\number\day, \number\year}

\def\tpoint{\tenpoint}
\def\npoint{\ninepoint}

\tpoint\rm

\hsize=5.5in
\hoffset=.5in

\def\tour{{\tit tour}}
\def\pushback{{\tit pushback}}
\outer\def\beginproclaim #1. {\medbreak
  \noindent{\bf#1.\enspace}\begingroup\sl}
\outer\def\endproclaim{\endgroup\par
  \ifdim\lastskip<\medskipamount \removelastskip\penalty55\medskip\fi}
\def\lvec#1{\overleftarrow{\strut#1}}
\def\rvec#1{\overrightarrow{\strut#1}}
\def\headmark{\ast}
\def\leftmark{\lvec\headmark}
\def\rightmark{\rvec\headmark}
\def\ul{\underline}

{\offinterlineskip

{\tpoint\rm\titless
\centerline{Counting Is Easy%
\footnote{$^{\dag}$}{{\npoint\rm
The work of the first author was
supported in part by the National Science Foundation under grant
MCS-8110430.}}}}

\vskip2pc

{\tpoint\rm
\centerline{\strut Joel I. Seiferas}
\centerline{Computer Science Department}
\centerline{University of Rochester}
\centerline{\strut Rochester, New York, U. S. A. 14627}}

\vskip1pc

{\tpoint\rm
\centerline{\strut Paul M. B. Vit\'anyi}
\centerline{Centre for Mathematics and Computer Science}
\centerline{P. O. Box 4079}
\centerline{\strut 1009 AB Amsterdam, The Netherlands}}

\vskip1pc

{\tpoint\rm
\centerline{\strut\date}}}

\vskip3pc

{\tpoint\rm
\narrower{\bf Abstract.}
For any fixed $k$, a remarkably simple single-tape Turing machine can 
simulate $k$ independent counters in real time.\par}

\vskip1pc

{\tpoint\rm
\narrower{\bf Categories and Subject Descriptors:}
F.1.1 [Computation by Abstract Devices]:  Models of
Computation---relations among models; bounded-action devices; E.2
[Data]:  Data Storage Representations---contiguous representations;
F.2.2 [Analysis of Algorithms and Problem Complexity]:  Nonnumerical
Algorithms and Problems---sequencing and scheduling; G.2.1 [Discrete
Mathematics]: Com\-bi\-na\-tor\-ics---combinatorial algorithms; F.2.2
[Analysis of Algorithms and Problem Complexity]:  Tradeoffs among
Complexity Measures\par}

\vskip1pc

{\tpoint\rm
\narrower{\bf General Terms:}
Theory, Algorithms, Design, Verification\par}

\vskip1pc

{\tpoint\rm
\narrower{\bf Additional Key Words and Phrases:}
Counter, abstract storage unit, counter machine, multicounter
machine, one-tape Turing machine, simulation between models,
real-time simulation, on-line simulation, oblivious simulation,
redundant number representation, signed-digit number representation,
recursion elimination\par}

\sectionskip\secbegin 1.  Introduction//
     In this paper we describe a remarkably simple real-time simulation, 
based on just five simple rewriting rules, of any fixed number $k$ of 
independent counters.  On a Turing machine with a single, binary work 
tape, the simulation runs in real time, handling an arbitrary counter 
command at each step.  The space used by the simulation can be held to 
$(k+\epsilon) \log_2 n$ bits for the first $n$ commands, for any 
specified $\epsilon > 0$.  Consequences and applications are discussed 
in [10--11], where the first single-tape, real-time simulation of 
multiple counters was reported.

     Informally, a {\demph counter\/} is a storage unit that maintains a 
single integer (initially $0$), incrementing it, decrementing it, or 
reporting its sign (positive, negative, or zero) on command.  Any 
automaton that responds to each successive command as a counter would is 
said to {\demph simulate\/} a counter.  (Only for a sign inquiry is the 
response of interest, of course.  And zeroness is the only real issue, 
since a simulator can readily use zero detection to keep track of 
positivity and negativity in finite-state control.)  To simulate $k$ 
independent counters, an automaton must respond to $3k$ commands:  
``increment counter number~$i$'', ``decrement counter number~$i$'', and 
``report the sign of counter number~$i$'' ($1 \le i \le k$).  If there 
is some fixed bound on the time needed by a simulator to respond to the 
successive commands it receives, then it simulates in {\demph real 
time}.

     Our real-time $k$-counter simulator will be a {\demph single-tape
Turing machine}.  Such an automaton consists of a finite-state control
unit with read-write access to an infinite but initially blank binary
storage tape (0 in every bit position).  Each next step is determined by
the current control state, the bit currently scanned by the read-write
head on the storage tape, and the most recently received input symbol
(in our case, the last command not yet responded to).  Each step can
involve any of the following actions:  a change to the bit scanned by
the head on the storage tape, a shift left or right by that head to an
adjacent bit position, emission of an output symbol (in our case, a
command response), and a state transition by the finite-state control
unit.

     An apparently stronger notion of real-time simulation would require
response to each successive command just {\emph one\/} step after
submission.  In the special case of {\emph counter\/} simulation,
however, {\emph any\/} real-time simulation actually does also yield a
real-time simulation in which the command-response delay is just $1$. 
(It is well known that a larger delay can be ``swept under the rug'' by
increasing the size of the alphabet used on the storage tape, but that
is not necessary in our case.)

     \proclaim Proposition.  If a single-tape Turing machine can
simulate $k$ counters in real time with command-response delay bound
$d$, then a similar single-tape Turing machine (still with only binary
tape alphabet) can do so with delay bound $1$.

     \pfbegin Proof.  The rough idea is for the delay-$1$ simulation to
use a delay-$d$ simulation to store an appropriate {\emph fraction\/} of
each of its counters' contents, and to maintain all the remainders in
finite-state control.

     More accurately and precisely, the delay-$1$ simulation can operate
in ``phases'' of $2kd$ steps, maintaining the following invariant from
phase to phase, for the absolute value $|c|$ of each count $c$:
                        $$|c| = c_0 + c_1(2kd),$$
where either
$$\displaylines{\hfill
        c_1>0 \quad\hbox{and}\quad 2kd \le c_0 \le 8kd,\hfill\cr
\noalign{\hbox{or}} \hfill
           c_1=0 \quad\hbox{and}\quad 0 \le c_0 \le 8kd,\hfill\cr}$$
and where $c_0$ and the signs of $c$ and $c_1$ are stored in finite-state
control, and $c_1$ is stored in the corresponding counter of the
delay-$d$ simulation.  The $2kd$ commands received in each phase can be
handled within finite-state control, increasing or decreasing each $c_0$ by at
most $2kd$.  Meanwhile, the $2kd$ steps are enough for one increment or
decrement of and one interrogation of each $c_1$.  In each case, the
simulation should {\emph increment\/} $c_1$, as part of a ``carry'' from
$c_0$, if $c_0>6kd$ held when the phase began; and it should {\emph
decrement\/} $c_1$, as part of a ``borrow'' for $c_0$, if $c_0<4kd$ held
when the phase began, unless $c_1$ was already zero.  For each count, if
$c_1$ was positive when the phase began, then $2kd \le c_0 \le 8kd$ will
hold when it ends.  If $c_1$ was zero when the phase began, however,
$c_0$ might ``underflow'' almost to~$-2kd$; but, in that case, $c_1$ will
remain zero, so that a sign change in finite-state control will suffice to
restore the invariant.  Finally, note that there will always be
enough information in finite-state control to determine whether a count is
currently zero:  Each count will be zero just when its $c_0$ is zero and
its $c_1$ was zero when the current phase began.\qed

\medskip

     Prior to the breakthrough in [10--11], there were at least
three weaker simulations in the literature.  M. Fischer and Rosenberg
[4] showed that the simulation is possible in the case that only {\emph
simultaneous\/} zeroness of the $k$ counters has to be reported.  P.
Fischer, Meyer, and Rosenberg [5] showed that a full simulation is
possible in cumulative {\emph linear\/} time (i.e., with {\emph
average\/} delay bounded by a constant, but with no fixed bound on the
delay for each individual command).  A while later, the latter authors
showed that {\emph four\/} Turing-machine tapes are as efficient as $k$
counters, for {\emph sequence generation\/} [6].  F\"urer's full
linear-time simulation [7] requires more than one tape, but two suffice
{\emph even if they are otherwise occupied}.

\sectionskip\secbegin 2.  A Peek at an Oblivious Solution//
     Using a straightforward unary, or ``tally'', notation, an automaton
with just one storage tape (i.e., a single-tape Turing machine) obviously
can simulate a single counter in real time.  An appropriate redundant 
variant of binary notation also suffices and requires much less space on
the storage tape [4].

     To simulate more than one counter in real time using a single tape 
is much harder.  For any $k$, in fact, it is hard to imagine how fewer 
than $k$ separate tapes can suffice to simulate $k$ counters in real 
time.  Since the contents of the counters to be simulated can fluctuate 
completely independently, we seem to be forced to consider simulations 
that actually handle the separate counters separately, say on $k$
separate ``tracks'' of the one available tape.  The problem is to assure
that the simulator's one tape head is always in the right place for
every one of these separate handlings, since the next command might be 
addressed to any of the simulated counters.

     Each ``separate handling'' above is essentially a real-time 
simulation of one counter.  The requirement that the tape head is always
in the right place can be formulated most clearly if our counters are
``enhanced'' to handle one additional command, a command to ``do
nothing''.  (Any efficient simulation of an unenhanced counter trivially
yields an efficient simulation of an enhanced one, anyway:  Simply
handle each ``do nothing'' as if it were an ``increment'' followed by a
``decrement''.)  Then we can view each command to a multiple-counter
storage unit as a tuple of commands, one to each separate counter.  What
we need, therefore, is a real-time single-counter simulation that is
``oblivious'' in the sense that neither its head position nor its times
of interaction with the outside world (to respond to commands and to
receive new ones) depend at all on the particular command sequence.  Our
real-time simulation of a $k$-counter storage unit is indeed based on
performing, on a separate track of the one available storage tape, just
such a simulation for each of the $k$ simultaneous command streams.

     In the rest of this section, without further motivation, we preview
the entire oblivious simulation of a single counter.  In the following
sections, on the other hand, we will return to an evolutionary top-down
development of the simulation, with each successive refinement motivated
by some outstanding inadequacy or loose end.  Having previewed the final
concrete result, the reader will better appreciate the direction and
progress of that evolution.

     For transparency, we actually implement our oblivious one-counter
simulation on a single-tape Turing machine model that is apparently
stronger than the one defined above.  The stronger model can write and
read symbols from some slightly larger alphabet on its storage tape, and
each next step can depend on, change, and shift among all the symbols in
some small {\emph neighborhood\/} of the head position on the storage
tape.  By coding in binary, and by conceding a somewhat larger (but still
fixed) bound on command-response delay time, however, we could
straightforwardly replace any such oblivious real-time simulator by an
oblivious one of the promised variety.

     Each nonblank storage tape symbol used by the simulator includes a
base symbol from the set $\{\hbox{-3}, \hbox{-2}, \hbox{-1}, 0, 1, 2, 3,
\ast\}$ and a left or right overarrow.  Optionally, it can also include
an underline and one or two primes.  The purpose of the base symbol
$\ast$ is to mark the position of the read-write head.  The initial
storage contents is treated as if it were
$$\cdots \> \lvec0 \> \lvec0 \> \lvec{0'} \> \rightmark \> \rvec0 \>
\rvec0 \> \rvec0 \> \cdots\,.$$

     With such a rich storage tape alphabet, our simulator will not have
to remember anything in finite-state control---a single state will
suffice.  Therefore, since even the head position will be implicit in
the contents of the storage tape, the transition rules will be just a
set of context-sensitive rules for rewriting the storage tape.  We
promised five such rules, but they are actually five entire {\emph
schemes\/}:
$$\vbox{\settabs
\+$a     \rvec b$&${}\leftmark c''\rvec d e$&${}\Rightarrow b   \leftmark  \rvec d c'' d$, propagating carry or borrow from $c$ to $b$\cr
\+$\hfill b$     &${}\leftmark c'$          &${}\Rightarrow {}  \rightmark b c$\cr
\medskip
\+$\hfill\rvec b$&${}\leftmark c$           &${}\Rightarrow c'  \rightmark \lvec b$, propagating into $b$, and then from $b$ to $c$\cr
\smallskip
\+$a     \lvec b$&${}\leftmark c$           &${}\Rightarrow ac''\rightmark \lvec b$, propagating into $b$, and then from $b$ to $a$\cr
\medskip
\+$\hfill b$     &${}\leftmark c''\lvec d$  &${}\Rightarrow b   \leftmark  \rvec d c'$, propagating from $d$ to $c$\cr
\smallskip
\+$\hfill b$     &${}\leftmark c''\rvec d e$&${}\Rightarrow b   \leftmark  \rvec d c'' e$, propagating from $d$ to $e$\cr}$$
Each of $a$, $b$, $c$, $d$, and $e$ can be any member of $\{\hbox{-3},
\hbox{-2}, \hbox{-1}, 0, 1, 2, 3\}$.  Except on the symbol with base
$c$, primes are not shown and are unchanged by the transitions. 
Similarly, arrows not shown are unchanged by the transitions.  The
mirror-image reflections of the rules describe the transitions when 
$\ast$ lies beneath a {\emph right\/} arrow; thus, for example, the very
first transition is according to the first of the five schemes, yielding
$$\cdots \> \lvec0 \> \lvec0 \> \lvec0 \> \rvec0 \> \leftmark \> \rvec0
\> \rvec0 \> \cdots\,.$$
Note that the rule for each next transition will be determined by the
number of primes on $c$ and the direction of the arrow over $b$ or
$d$, and that the symbols playing these roles will be determined by
the direction of the arrow over $\ast$.  It remains only to give the
rules for information ``propagation'', for maintenance of the
underlines (not shown in the rule schemes), and for generation of
responses to the commands.

     ``Propagation from $b$ to $c$'' is essentially a ``carry'' or
``borrow'' operation:  If $b$ is $3$, then reduce it by $4$ (to
$\hbox{-1}$) and add $1$ to $c$.  If $b$ is $\hbox{-3}$, then increase
it by $4$ (to $1$) and subtract $1$ from $c$.  If either of these
actions changes $c$ {\emph to\/} $0$, and $c$ was not underlined, then
remove the underline from $b$; and, if either action changes $c$ {\emph
from\/} $0$, and $b$ was not underlined, then add an underline to $b$. 
Leave all other underlining unchanged.

     ``Propagation into $b$'' depends on the next input command.  On a
command to increment or decrement the counter, $b$ is incremented or
decremented accordingly.  The result is a count of zero if and only if
the resulting $b$ is $0$, without an underline.

     The delay between the handling of successive input commands is at
most three steps, counts of zero are detected correctly, and no base
symbol is ever required to overflow past~$3$ or to underflow past
$\hbox{-3}$, although these facts are not at all clear from just the
rules.  It {\emph is\/} clear from the rules that the simulation is
both deterministic and oblivious.

     As an example, suppose every command is to {\emph increment\/} the
counter.  Then the results of the first six transitions are as follows:
$$\displaylines{
\cdots \> \lvec0 \> \lvec0 \> \rvec0 \> \leftmark \> \rvec0 \> \rvec0 \>
\rvec0 \> \rvec0 \> \cdots\qquad\hbox{(by rule 1),}\cr
\cdots \> \lvec0 \> \lvec0 \> \rvec{0'} \> \rightmark \> \lvec1 \> \rvec0
\> \rvec0 \> \rvec0 \> \cdots\qquad\hbox{(by rule 2),}\cr
\cdots \> \lvec0 \> \lvec0 \> \rvec0 \> \lvec1 \> \leftmark \> \rvec0 \>
\rvec0 \> \rvec0 \> \cdots\qquad\hbox{(by rule 1),}\cr
\cdots \> \lvec0 \> \lvec0 \> \rvec0 \> \rvec{0''} \> \rightmark \>
\lvec2 \> \rvec0 \> \rvec0 \> \cdots\qquad\hbox{(by rule 3),}\cr
\cdots \> \lvec0 \> \lvec0 \> \rvec{0'} \> \lvec0 \> \rightmark \>
\lvec2 \> \rvec0 \> \rvec0 \> \cdots\qquad\hbox{(by rule 4),}\cr
\cdots \> \lvec0 \> \lvec0 \> \rvec{0'} \> \ul{\rvec{\hbox{-1}}} \>
\leftmark \> \lvec{1'} \> \rvec0 \> \rvec0 \> \cdots\qquad\hbox{(by
rule 2).}\cr
}$$
Continuing in this way, the result of the first 2{,}980{,}000
transitions, including the execution of 1{,}191{,}993 commands to
increment the counter, is
$$\cdots \> \lvec0 \> \lvec0 \> \rvec{0'} \> \rvec{0'} \> \ul{\rvec0} \>
\rvec{1''} \> \ul{\rvec{0'}} \> \ul{\rvec2} \> \ul{\rvec{\hbox{-1}''}}
\> \rightmark \> \ul{\lvec1} \> \ul{\rvec{1'}} \> \ul{\lvec0} \>
\ul{\lvec{2''}} \> \ul{\rvec{\hbox{-1}}} \> \ul{\lvec1} \> \lvec{0'} \>
\lvec{0'} \> \rvec0 \rvec0 \> \cdots\,.$$
(For now, this should seem pretty obscure as a representation for
1{,}191{,}993.  It will turn out that the base symbols are a scrambled
radix-$4$ representation for that number: 
$$\big(1021(\hbox{-1})001(\hbox{-1})21\big)_4 = \big(1191993\big)_{10}.$$
The unscrambled order is implicit in the arrows and primes.  The
underlining indicates which radix-$4$ digits are significant, except that
the {\emph leading\/} significant digit is {\emph not\/} underlined. 
(In a radix number, a digit is {\demph significant\/} as long as it is
not a leading $0$.))

\sectionskip\secbegin 3.  Oblivious Counting//
     There is a relatively familiar technique that makes it possible to
maintain a counter obliviously in real time {\emph if\/} the oblivious
order of position access can be nonsequential.  The oblivious version
[9] of the classical two-tape simulation [8] of multiple Turing-machine
tapes is based implicitly on the technique.  The technique involves a
liberalization of ordinary fixed-radix notation, allowing an expanded
range of ``signed digits'' in each position [2, 1].  This, in turn, 
allows some choice on numbers' representations and some optional delay 
in carry propagation.  To maintain such a representation as the 
represented number is incremented and decremented, we need only visit 
the various positions often enough to avoid overflow and underflow.  The
following two requirements, which are oblivious to the particular 
sequence of commands, are sufficient for such a scheme to be able to 
handle commands in real time:

\medskip
     \item{1.}  There is a chance (``$0$-opportunity'') to propagate 
information (increments and decrements) into position $0$ at least once 
every  $O(1)$  steps.

     \item{2.}  There is a chance (``$(i+1)$-opportunity'') to 
propagate information (carries and borrows) from position $i$ into 
position $i+1$ at least once every  $O(1)$  times there is an 
$i$-opportunity.
\medskip

\noindent These requirements are met, for example, by a schedule that
provides a $0$-opportunity every other step, a $1$-opportunity every
other remaining step, a $2$-opportunity every other still remaining
step, etc.:
     $$0\,1\,0\,2\,0\,1\,0\,3\,0\,1\,0\,2\,0\,1\,0\,4\,
       0\,1\,0\,2\,0\,1\,0\,3\,0\,1\,0\,2\,0\,1\,0\,5\,
       0\,1\,0\,2\,0\,1\,0\,3\,0\,1\,0\,2\,0\,1\,0\,4\,
       0\,1\,0\,2\ldots\,{}.$$
This is the sequence of carry propagation distances when we count in
binary, so let us call it the {\demph binary carry schedule}.

     To see that the requirements suffice, consider using a radix $r$ 
that is large compared to the constants (``$O(1)$'') with which the 
requirements are satisfied.  Symmetrically allow as ``digits'' all 
integers~$d$  satisfying  $-r < d < +r$.  (For our ultimate use, the 
radix  $r = 4$  will be large enough; this explains the use of the digit
set $\{\hbox{-3}, \hbox{-2}, \hbox{-1}, 0, 1, 2, 3\}$ in Section~2's
preview of the simulation.)  As suggested in Section 2, maintain an
underline beneath each significant digit except for the leading one. 
Propagate information from position $i$ to position $i+1$ according to
the following simple rules:
$$\vbox{\+``Carry'' if the digit is greater than $r/2$.\cr
        \+``Borrow'' if the digit is less than $-r/2$.\cr
        \+Do nothing if the digit is bounded by $r/2$ in absolute value.\cr}$$
(To ``carry'', reduce the digit in position~$i$ by~$r$, and increment 
the digit in position~$i+1$ by~$1$.  To ``borrow'', reduce the digit in 
position~$i+1$ by~$1$, and increment the digit in position~$i$ by~$r$.) 
By induction, the properties of the maintenance schedule assure that no 
digit will have to exceed $r-1$ in absolute value.  As a consequence, 
the only digit that might change from zero to nonzero, or vice versa, is
at position $i+1$ above, so that only the underlining at position $i$
might have to change, and so that correct underlining can be maintained
without any additional access to the digits of the counter.  As another 
consequence, the leading significant digit (if there is one) will
always correctly indicate the sign of the entire count, so that the
count will be $0$ only when the frequently observed digit at position
$0$ is a $0$ with no underline.

\sectionskip\secbegin 4.  Permutation for Sequential Access//
     With only {\emph sequential\/} access, it seems impossible to visit 
the positions of a radix number according to the scheduling requirements 
above.  The first requirement keeps us close to the low-order digit, 
while the second requirement draws us to arbitrarily-high-order digits. 
This intuition is wrong, however; even with the strictly sequential 
access available on a single Turing machine tape, we {\emph can\/} satisfy 
the requirements.  The trick is to maintain, on the main track of the 
tape, an appropriate, dynamically (but obliviously) changing {\emph 
permutation\/} of the radix positions.  We turn now to a top-down 
development and implementation of a suitable permutation procedure.

     Since the permutation procedure will be oblivious to the actual 
contents of the radix positions, and since position numbers will 
greatly clarify the permutation being performed, we will speak as if we 
are permuting the position numbers themselves.  It is important to 
remember, however, that it will be impossible with any finite tape 
alphabet for our simulator to maintain these unbounded position numbers 
on its tape without using too much space and time.  To recognize what 
may be obvious from the position numbers, the ultimate simulator will
have to cleverly maintain auxiliary markers from some finite alphabet
(primes, double primes, and overarrows in the simulation we describe) on
an auxiliary track of its tape.

     Consider the problem of visiting the positions of a radix number
according to the binary carry schedule.  The key to the schedule is that
it brackets each visit to position $i+1$ by full ``tours'' of positions
$0$,~\dots,~$i$, denoted by $\tour(i)$:
$$\vbox{\settabs\+\quad&visit $i+1$\qquad\qquad&\quad&\cr
                \+$\tour(i+1)$:&               &$\tour(0)$:\cr
                     \+&$\tour(i)$             &     &visit $0$\cr
                     \+&visit $i+1$\cr
                     \+&$\tour(i)$\cr}$$
Noting that appending ``visit $i+1$; $\tour(i)$'' onto the end of 
$\tour(i)$ always gives $\tour(i+1)$, we see that $\tour(\infty)$ makes 
sense:
$$\vbox{\settabs\+\quad&visit $3$; $\tour(2)$&\cr
                \+$\tour(\infty)$:\cr
                     \+&visit $0$\cr
                     \+&visit $1$; $\tour(0)$\cr
                     \+&visit $2$; $\tour(1)$\cr
                     \+&visit $3$; $\tour(2)$\cr
                     \+&\hfill$\vdots$\hfill&\cr}$$
In fact $\tour(\infty)$ is precisely the entire binary carry schedule.

     For a Turing machine implementation of all this touring, we must 
permute to keep the head, represented by $\headmark$ as in Section 2,
always near position number $0$.  Thus we might try the permutational
side effect
     $$\tour(i)\colon  {}\headmark 012 \ldots i(i+1) \Rightarrow 
                                       i \ldots 210 \headmark (i+1)$$
as preparation for the first visit to  $i + 1$.  But then $\tour(i)$ (or 
even its symmetric mirror image) would no longer complete the desired 
analogous preparation (i.e., $\tour(i+1)$) for the first visit to $i + 
2$.  With the latter goal in mind, we are led to push position  $i + 1$ 
left during the second (mirror-image) iteration of $\tour(i)$ and to 
introduce into $tour(i+1)$ a third iteration of $\tour(i)$, to get back 
to position $i+2$.  This way, the permutational side effect of 
$tour(i+1)$ is from ${}\headmark 012 \ldots i(i+1)$ initially, to $i 
\ldots 210 \headmark (i+1)$ after the first iteration of $tour(i)$, to 
$(i+1) \headmark 012 \ldots i$ after the second iteration, finally to 
$(i+1)i \ldots 210 \headmark{}$ after the third iteration, as desired.  
This leads us to refine our terminology, in order to reflect the two 
variants of {\demph $i$-tour\/} ($\tour(i)$  above):
$$\vbox{\settabs
    \+$\tour(i,+)$:       &$j \headmark 012 \ldots i$&${}\Rightarrow 
                                              i \ldots 210 \headmark j$\quad &{\npoint\rm(mirror image)}\cr
    \smallskip
    \+$\tour(i,-)$:  &\hfill   $\headmark\,012 \ldots i$&${}\Rightarrow  
                                              i \ldots 210\,\headmark$\cr
    \+$\tour(i,-)$:  &\hfill   $i \ldots 210\,\headmark$&${}\Rightarrow  
                                              \headmark\,012 \ldots i$\hfill &{\npoint\rm(mirror image)}\cr
    \medskip
    \+$\tour(i,+)$:       &$j \headmark 012 \ldots i$&${}\Rightarrow 
                                              i \ldots 210 \headmark j$\cr
    \+$\tour(i,+)$:       &$i \ldots 210 \headmark j$&${}\Rightarrow 
                                              j \headmark 012 \ldots i$      &{\npoint\rm(mirror image)}\cr
    \smallskip}$$
We will refer to these variants as {\demph negative} $i$-tours and 
{\demph positive} $i$-tours, respectively, depending on whether some 
position $j$ {\emph is\/} or {\emph is not\/} being ``transported''. 
Note that we do not distinguish notationally between a tour and its
mirror image, since only one of the two can be applicable at a time,
depending on the current location of position $0$. Similarly, we do not
incorporate into the notation the position $j$ being transported by a
positive tour, since there is never any choice.

     Suppressing explicit visits now (since convenient $i$-opportunities 
will arise at a different point in our scheme, and since the visits do 
not affect the actual permutation process anyway), we arrive at the 
following mutually recursive implementations for our evolving tours (the 
program locations are labeled (a) through (e) for later reference):
$$\vbox{\settabs\+\quad&(a) &$\tour(i,+)$\qquad&                                            \cr
\smallskip
                \+$\tour(i+1,-)$:&&            &{\npoint\rm(start with ${}\headmark 012 \ldots i(i+1)$)}\cr
                \+     &    &$\tour(i,-)$      &{\npoint\rm(permute to $i \ldots 210 \headmark (i+1)$)} \cr
                \+     &(a) &$\tour(i,+)$      &{\npoint\rm(permute to $(i+1) \headmark 012 \ldots i$)} \cr
                \+     &(b) &$\tour(i,-)$      &{\npoint\rm(permute to $(i+1)i \ldots 210 \headmark{}$)}\cr
\medskip
          \settabs\+\quad&(d) &$\tour(i,+)$\qquad&                                            \cr
                  \+$\tour(i+1,+)$:&&            &{\npoint\rm(start with $j \headmark 012 \ldots i(i+1)$)}\cr
                  \+     &    &$\tour(i,+)$      &{\npoint\rm(permute to $i \ldots 210 \headmark j(i+1)$)}\cr
                  \+     &(c) &$\pushback$       &{\npoint\rm(permute to $i \ldots 210 \headmark (i+1)j$)}\cr
                  \+     &(d) &$\tour(i,+)$      &{\npoint\rm(permute to $(i+1) \headmark 012 \ldots ij$)}\cr
                  \+     &(e) &$\tour(i,-)$      &{\npoint\rm(permute to $(i+1)i \ldots 210 \headmark j$)}\cr
\smallskip}$$
The recursive strategy for $tour(i+1,-)$ is as described previously, but 
the strategy for $tour(i+1,+)$ is new.  Note that the latter requires a 
new permutation step, called a {\demph pushback\/}, to push the nonzero 
position currently adjacent to the head beyond the next adjacent 
position.  Finally, since appending 
             $$\hbox{(a) $\tour(i,+)$; (b)  $\tour(i,-)$}$$
onto the end of $\tour(i,-)$ always gives $\tour(i+1,-)$, we again have
a well-defined infinite limit:
$$\vbox{
\smallskip
          \settabs\+\quad&(a) $\tour(2,+)$; (b) &$\tour(2,-)$\qquad & \cr
                  \+$\tour(\infty,-)$:&         &                   &{\npoint\rm(start with ${}\headmark 012345 \ldots$)} \cr
                  \+     &                      &$\tour(0,-)$       &{\npoint\rm(permute to $0 \headmark 12345 \ldots$)}  \cr
                  \+     &(a) $\tour(0,+)$; (b) &$\tour(0,-)$       &{\npoint\rm(permute to $10 \headmark 2345 \ldots$)}  \cr
                  \+     &(a) $\tour(1,+)$; (b) &$\tour(1,-)$       &{\npoint\rm(permute to $210 \headmark 345 \ldots$)}  \cr
                  \+     &(a) $\tour(2,+)$; (b) &$\tour(2,-)$       &{\npoint\rm(permute to $3210 \headmark 45 \ldots$)}  \cr
          \settabs\+\quad&(a) $\tour(2,+)$; (b)  $\tour(2,-)$&\qquad&{\npoint\rm(start with ${}\headmark 012345 \ldots$)}&\cr
                  \+     &\hfill$\vdots$\hfill               &      &\hfill$\vdots$\hfill                                   &\cr
\smallskip}$$
It is $tour(\infty,-)$ that we actually implement.

\sectionskip\secbegin 5.  Recursion Elimination//
     By induction, the entire permutation process $tour(\infty,-)$
involves just three, symmetric pairs of atomic moves:
$$\vbox{\+\cleartabs
$\tour(0,-)$:  &$\headmark\,0 \Rightarrow 0\,\headmark$\qquad\qquad
$\tour(0,+)$:  &$j \headmark 0 \Rightarrow 0 \headmark j$\qquad\qquad
$\pushback$:   &$0 \headmark ji \Rightarrow 0 \headmark ij$\cr
\+             &$0\,\headmark \Rightarrow \headmark\,0$
               &$0 \headmark j \Rightarrow j \headmark 0$
               &$ij \headmark 0 \Rightarrow ji \headmark 0$\cr}$$
Our simulator will have to determine which of these local permutations
to perform at each step.  The problem is analogous to the derivation
of a nonrecursive solution to the ``Towers of Hanoi'' problem from
the more obvious recursive solution [3].  In this section we solve
the problem by adding a small number of carefully chosen notations to
the symbols being permuted.

     Because position $0$ will always be to the immediate left or right 
of the head, the simulator will be able to maintain the correct current 
direction to position $0$ in finite-state control, narrowing the possibilities 
to just one atomic move from each pair above.  The remaining problem is 
to determine whether the next step should be a negative $0$-tour, a 
positive $0$-tour, or a pushback.

     \proclaim Observation 1.  For every $i$, the respective first moves 
of  $\tour(i,-)$  and  $\tour(i,+)$  are  $\tour(0,-)$  and  
$\tour(0,+)$.

     Except for the initial situation, when  $\tour(0,-)$  is required
explicitly, program locations (a) through (e) account for all situations.
By Observation 1, it will suffice always to know whether the next move 
starts a negative tour (program locations (b), (e)), starts a positive 
tour (program locations (a), (d)), or is a pushback (program location 
(c)).  A good clue would be the largest action that the {\emph
previous\/} move {\emph ended\/}; this clue is not readily available,
however, since negative $(i+1)$-tours and positive $(i+1)$-tours both
end with the same move, a negative $0$-tour.

     \proclaim Observation 2.  A positive tour ends with the head 
adjacent to the transported position~$j$.

     \proclaim Observation 3.  By induction, at no time properly within 
a tour is the head adjacent to a position not explicitly involved in the 
tour.  (The positions explicitly involved in  $\tour(i,-)$  are $0$ 
through $i$, and the ones explicitly involved in  $\tour(i,+)$  are 
these and also the transported position $j$.)

     \proclaim Corollary.  The position $j$ that gets pushed back at the 
outermost level of a positive tour will next be adjacent to the head at 
the end of that positive tour.

     This last corollary presents an opportunity to recognize the end of 
a positive tour:  The head can leave a ``message'' attached to the 
position that gets pushed back, indicating that a positive tour is in 
progress.  (In our ultimate implementation, the messages will be single 
and double primes on symbols.)  Consequently, the simulator will be able 
to recognize when a positive tour is ending, at which time it can delete 
the message (remove the single or double prime).  (In the special case 
of the one-move positive tour $\tour(0,+)$, there is no pushback; in 
this case, for uniformity, the same sort of message can be attached to 
the relevant position $j$, in the one move that does take place.)  The 
absence of such a message, therefore, will surely indicate program 
location (a) or (d) and hence that the next move should be  
$\tour(0,+)$.  In the {\emph presence\/} of such a message, however, it 
still remains to distinguish program location (c) (which is followed by 
a pushback) from program locations (b) and (e) (which are followed by  
$\tour(0,-)$).  For this purpose, we introduce an auxiliary distinction 
between two varieties of positive tour, a distinction that we will try 
to record as part of the message corresponding to the positive tour.  
The distinction is simply $j = i+1$ versus $j > i+1$:
$$\vbox{\settabs
     \+$\tour''(i,+)$:  &$       (i+1) \headmark 012 \ldots i$&${}\Rightarrow
                                  i \ldots 210 \headmark j$\qquad ($j > i+1$)\cr
     \smallskip
     \+$\tour'(i,+)$:   &$\hfill (i+1) \headmark 012 \ldots i$&${}\Rightarrow
                                  i \ldots 210 \headmark (i+1)$\cr
     \+$\tour'(i,+)$:   &$\hfill i \ldots 210 \headmark (i+1)$&${}\Rightarrow
                                  (i+1) \headmark 012 \ldots i$\cr
     \medskip
     \+$\tour''(i,+)$:  &$\hfill     j \headmark 012 \ldots i$&${}\Rightarrow
                                  i \ldots 210 \headmark j$\qquad ($j > i+1$)\cr
     \+$\tour''(i,+)$:  &$\hfill     i \ldots 210 \headmark j$&${}\Rightarrow
                                  j \headmark 012 \ldots i$\qquad ($j > i+1$)\cr
     \smallskip}$$
In the correspondingly revised recursion, doubly primed positive tours 
are needed only for the first subtour at the outermost level of each 
positive tour.  Because different messages have to be left, we begin now 
to distinguish between singly and doubly primed pushbacks.  For use in 
our analysis, we add the recursion level of a pushback to the notation, 
even though it is not algorithmically significant.
$$\vbox{\settabs\+\quad&(a) &$\tour'(i,+)$\qquad&                                            \cr
\smallskip
                \+$\tour(i+1,-)$:&&             &{\npoint\rm(start with ${}\headmark 012 \ldots i(i+1)$)}\cr
                \+     &    &$\tour(i,-)$       &{\npoint\rm(permute to $i \ldots 210 \headmark (i+1)$)} \cr
                \+     &(a) &$\tour'(i,+)$      &{\npoint\rm(permute to $(i+1) \headmark 012 \ldots i$)} \cr
                \+     &(b) &$\tour(i,-)$       &{\npoint\rm(permute to $(i+1)i \ldots 210 \headmark{}$)}\cr
\medskip
          \settabs\+\quad&(d) &$\pushback'(i+1)$\qquad&                                                               \cr
                  \+$\tour'(i+1,+)$:&&                &{\npoint\rm(start with $(i+2) \headmark 012 \ldots i(i+1)$)}\cr
                  \+     &    &$\tour''(i,+)$         &{\npoint\rm(permute to $i \ldots 210 \headmark (i+2)(i+1)$)}\cr
                  \+     &(c) &$\pushback'(i+1)$      &{\npoint\rm(permute to $i \ldots 210 \headmark (i+1)(i+2)$)}\cr
                  \+     &(d) &$\tour'(i,+)$          &{\npoint\rm(permute to $(i+1) \headmark 012 \ldots i(i+2)$)}\cr
                  \+     &(e) &$\tour(i,-)$           &{\npoint\rm(permute to $(i+1)i \ldots 210 \headmark (i+2)$)}\cr
\medskip
          \settabs\+\quad&(d) &$\pushback''(i+1)$\qquad&                                                           \cr
                  \+$\tour''(i+1,+)$:&&                &{\npoint\rm(start with $j \headmark 012 \ldots i(i+1)$)}\cr
                  \+     &    &$\tour''(i,+)$          &{\npoint\rm(permute to $i \ldots 210 \headmark j(i+1)$)}\cr
                  \+     &(c) &$\pushback''(i+1)$      &{\npoint\rm(permute to $i \ldots 210 \headmark (i+1)j$)}\cr
                  \+     &(d) &$\tour'(i,+)$           &{\npoint\rm(permute to $(i+1) \headmark 012 \ldots ij$)}\cr
                  \+     &(e) &$\tour(i,-)$            &{\npoint\rm(permute to $(i+1)i \ldots 210 \headmark j$)}\cr
\medskip
          \settabs\+\quad&(a) $\tour(2,+)$; (b) &$\tour(2,-)$\qquad &                                                         \cr
                  \+$\tour(\infty,-)$:&          &                   &{\npoint\rm(start with ${}\headmark 012345 \ldots$)} \cr
                  \+     &                       &$\tour(0,-)$       &{\npoint\rm(permute to $0 \headmark 12345 \ldots$)}  \cr
                  \+     &(a) $\tour'(0,+)$; (b) &$\tour(0,-)$       &{\npoint\rm(permute to $10 \headmark 2345 \ldots$)}  \cr
                  \+     &(a) $\tour'(1,+)$; (b) &$\tour(1,-)$       &{\npoint\rm(permute to $210 \headmark 345 \ldots$)}  \cr
                  \+     &(a) $\tour'(2,+)$; (b) &$\tour(2,-)$       &{\npoint\rm(permute to $3210 \headmark 45 \ldots$)}  \cr
          \settabs\+\quad&(a) $\tour'(2,+)$; (b)  $\tour(2,-)$&\qquad&{\npoint\rm(start with ${}\headmark 012345 \ldots$)}&\cr
                  \+     &\hfill$\vdots$\hfill               &       &\hfill$\vdots$\hfill                                   &\cr
\smallskip}$$
As desired, now, the end of the doubly primed variety of positive tour 
will indicate program location (c), and the end of the singly primed 
variety will indicate program location (b)~or~(e).

     It remains to find a way to recognize which variety of positive 
tour each pushback is a top-level part of, and which is the variety of 
each positive $0$-tour, so that the right messages (single or double 
prime, corresponding to the singly or doubly primed variety of pushback
or positive tour) can be recorded.  For these purposes, we will maintain
with each position the direction in the current permutation to its
successor. (This is the purpose of the overarrows.)  When we summarize
in Section 7, we will indicate how to keep this information up to date. 
To see that this directional information will help, we need one more
inductive observation:

     \proclaim Observation 4.  In each invocation of  $\tour''(i,+)$ 
(only two possibilities above), the first uninvolved position initially 
beyond position $i$ is position $i+1$.  (In either case, the initial 
permutation will include $j \headmark 012 \ldots i(i+1)$ or its mirror 
image.)

\noindent In all our invocations of  $\tour''(i+1,+)$, therefore, the 
first uninvolved position initially beyond position $i+1$ will be 
position $i+2$, so that the precondition for  $\pushback''(i+1)$  will 
always be  $0 \headmark j(i+1)(i+2)$  (or its mirror image).  The 
precondition for $\pushback'(i+1)$, on the other hand, will always be $0 
\headmark (i+2)(i+1)$ (or its mirror image).  The distinction can be 
recognized from the directional information for position $i+1$.  
Similarly, the precondition for  $\tour''(0,+)$ will always be  $10 
\headmark j$ (or its mirror image), while the precondition for 
$\tour'(0,+)$ will always be  $0 \headmark 1$ (or its mirror image), a 
distinction that can be recognized from the directional information for 
position $0$.

     In summary, here are suitable specifications for the evolved
versions of all the tours and pushbacks (except for mirror images),
now showing single- and double-prime messages (but not showing
overarrows, since we are still showing explicit position numbers):
$$\vbox{\settabs
    \+$\pushback''(i)$:  &$j \headmark 012 \ldots i(i+1)$&${}\Rightarrow 
                           i \ldots 210 \headmark j''(i+1)$\cr
    \smallskip
    \+$\tour(i,-)$:      &\hfill ${}' \headmark 012 \ldots i$&${}\Rightarrow  
                           i \ldots 210 \headmark {}$\cr
    \medskip
    \+$\tour'(i,+)$:     &$\hfill (i+1) \headmark 012 \ldots i$&${}\Rightarrow
                           i \ldots 210 \headmark (i+1)'$\cr
    \+$\tour''(i,+)$:    &$\hfill j \headmark 012 \ldots i(i+1)$&${}\Rightarrow
                           i \ldots 210 \headmark j''(i+1)$\qquad ($j > i+1$)\cr
    \medskip
    \+$\pushback'(i)$:   &$\hfill 0 \headmark (i+1)''i$&${}\Rightarrow
                           0 \headmark i(i+1)'$\cr
    \+$\pushback''(i)$:  &$\hfill 0 \headmark j''i(i+1)$&${}\Rightarrow
                           0 \headmark ij''(i+1)$\qquad ($j > i+1$)\cr
    \smallskip}$$
It is easy to check inductively that the recursive implementations do
maintain the specifications:
$$\vbox{\settabs\+\quad&$\tour'(i,+)$\qquad&                                                              \cr
\smallskip
                \+$\tour(i+1,-)$:&         &{\npoint\rm(start with ${}' \headmark 012 \ldots i(i+1)$)} \cr
                \+     &$\tour(i,-)$       &{\npoint\rm(permute to $i \ldots 210 \headmark (i+1)$)}    \cr
                \+     &$\tour'(i,+)$      &{\npoint\rm(permute to $(i+1)' \headmark 012 \ldots i$)}   \cr
                \+     &$\tour(i,-)$       &{\npoint\rm(permute to $(i+1)i \ldots 210 \headmark{}$)}   \cr
\medskip
          \settabs\+\quad&$\pushback'(i+1)$\qquad&                                                                 \cr
                  \+$\tour'(i+1,+)$:&            &{\npoint\rm(start with $(i+2) \headmark 012 \ldots i(i+1)$)}  \cr
                  \+     &$\tour''(i,+)$         &{\npoint\rm(permute to $i \ldots 210 \headmark (i+2)''(i+1)$)}\cr
                  \+     &$\pushback'(i+1)$      &{\npoint\rm(permute to $i \ldots 210 \headmark (i+1)(i+2)'$)} \cr
                  \+     &$\tour'(i,+)$          &{\npoint\rm(permute to $(i+1)' \headmark 012 \ldots i(i+2)'$)}\cr
                  \+     &$\tour(i,-)$           &{\npoint\rm(permute to $(i+1)i \ldots 210 \headmark (i+2)'$)} \cr
\medskip
          \settabs\+\quad&$\pushback''(i+1)$\qquad&                                                                   \cr
                  \+$\tour''(i+1,+)$:&            &{\npoint\rm(start with $j \headmark 012 \ldots i(i+1)(i+2)$)}   \cr
                  \+     &$\tour''(i,+)$          &{\npoint\rm(permute to $i \ldots 210 \headmark j''(i+1)(i+2)$)} \cr
                  \+     &$\pushback''(i+1)$      &{\npoint\rm(permute to $i \ldots 210 \headmark (i+1)j''(i+2)$)} \cr
                  \+     &$\tour'(i,+)$           &{\npoint\rm(permute to $(i+1)' \headmark 012 \ldots ij''(i+2)$)}\cr
                  \+     &$\tour(i,-)$            &{\npoint\rm(permute to $(i+1)i \ldots 210 \headmark j''(i+2)$)} \cr
\medskip
          \settabs\+\quad&$\tour(2,+)$; $\,$&$\tour(2,-)$\qquad &                                                         \cr
                  \+$\tour(\infty,-)$:&     &                   &{\npoint\rm(start with ${}'\headmark 012345 \ldots$)} \cr
                  \+     &                  &$\tour(0,-)$       &{\npoint\rm(permute to $0 \headmark 12345 \ldots$)}   \cr
                  \+     &$\tour'(0,+)$;    &$\tour(0,-)$       &{\npoint\rm(permute to $10 \headmark 2345 \ldots$)}   \cr
                  \+     &$\tour'(1,+)$;    &$\tour(1,-)$       &{\npoint\rm(permute to $210 \headmark 345 \ldots$)}   \cr
                  \+     &$\tour'(2,+)$;    &$\tour(2,-)$       &{\npoint\rm(permute to $3210 \headmark 45 \ldots$)}   \cr
          \settabs\+\quad&$\tour'(2,+)$;     $\tour(2,-)$&\qquad&{\npoint\rm(start with ${}\headmark 012345 \ldots$)} &\cr
                  \+     &\hfill$\vdots$\hfill                  &       &\hfill$\vdots$\hfill                                    &\cr
\smallskip}$$

\sectionskip\secbegin 6.  Opportunities To Carry and To Borrow//
     We see from the above preconditions for $\pushback'(i+1)$ and 
$\pushback''(i+1)$ (the ``$(i+1)$-pushbacks'') that these operations can
serve as $(i+2)$-opportunities.  Similarly, $\tour'(0,+)$ and 
$\tour''(0,+)$ (the ``positive $0$-tours'') can serve as 
$1$-opportunities.  Since the head is always adjacent to position $0$,
every step is a good time to propagate increments and decrements into
position~$0$; if we designate only the positive $0$-tours as
$0$-opportunities, however, we will ultimately be able to choose a
slightly smaller radix for our notation.

     \proclaim Observation 5.  If we omit $j$-pushbacks for $j > i$, 
then  $\tour(\infty,-)$ is an infinite concatenation of negative and
positive $i$-tours, the first of which is negative, the second of which
is positive, and no three consecutive of which are all negative or all
positive.  (To see the last part, make the analogous observation
by induction on $i' \ge i$ for each negative and positive $i'$-tour, and
finally note that $\tour(\infty,-)$ is the limit of the negative tours.)

     \beginproclaim Corollary.  In $\tour(\infty,-)$, our information 
propagation requirements are satisfied with respective constants $3$ 
and $4$:

\medskip
     \item{1.}  There is a $0$-opportunity at least once every three 
steps.

     \item{2.}  There is a $1$-opportunity every time there is a 
$0$-opportunity.

     \item{3.}  There are exactly two $(i+1)$-opportunities before the
first $(i+2)$-opportunity, and at most four $(i+1)$-opportunities
between $(i+2)$-opportunities.
\medskip\endproclaim

     \pfbegin Proof of third part.  The $(i+1)$-opportunities are 
distributed one per positive $i$-tour.  Using Observation 5 to focus on 
$(i+1)$-tours, therefore, we see that each negative tour presents one 
$(i+1)$-opportunity and no $(i+2)$-opportunity, and that each positive 
tour presents one $(i+1)$-opportunity before its one $(i+2)$-opportunity 
and one after.  The two initial $(i+1)$-opportunities come from the 
initial negative and positive tours, and the maximum of four intervening 
$(i+1)$-opportunities arise when a consecutive pair of negative tours is 
bracketed by positive tours.\qed

\medskip\noindent Since $5 + 4 \le 10 - 1$, it follows that $r = 10$ 
will be a large enough radix.  The more careful analysis in Section 8
reveals that even $r=4$ will be large enough.

\sectionskip\secbegin 7.  Formal Summary//
     In Section 5 we showed how to annotate the symbols being
permuted in the recursively defined $\tour(\infty,-)$ in such a way
that the very same permutation can be carried out nonrecursively by
a deterministic single-tape Turing machine, based entirely on local cues.
In Section 6 we observed that the same annotations provide sufficient
cues for adequate opportunities to perform the increments, decrements,
carries, and borrows required for our real-time simulation of a
counter.  In this section we finally relate all this to the few
simple rules previewed in Section 2.

     For transparency, we will summarize the rules we have derived in 
three increasingly formal stages.  In increasing order of difficulty, 
the four main cases are the first move, the case when a single-prime 
message is received, the case when no message is received, and the case 
when a double-prime message is received.  The first move is always  
$\tour(0,-)$.  When a single-prime message is received, $\tour(0,-)$ is 
again the correct move.  When no message is received, the correct move is 
either $\tour'(0,+)$ or $\tour''(0,+)$, depending on whether position 
$1$ is adjacent to the head or beyond position $0$; either way, a carry 
or borrow can be propagated as described above, and an indicative 
message should be left with the transported position.  When a 
double-prime message is received, the correct move is a singly or doubly 
primed pushback, depending on directional information near the head as 
described above; either way, a carry or borrow can be propagated as 
described above, and an indicative message should be left with the 
position that is pushed back.

     In the second stage, we reformulate our summary via formal rules in 
terms of position numbers.  For the messages corresponding to completion 
of singly and doubly primed positive tours, we use single and double 
primes on the position numbers.  Except in the case of the special rule 
for the very first move (${}\headmark 0 \Rightarrow 0 \headmark{}$), the 
mirror image of each rule is also a rule; so we will list only rules 
with position $0$ initially to the {\emph left\/} of the head.  Only
on the {\emph other\/} side of the head do we show primes explicitly,
since these primes constitute the message being received.
$$\vbox{\settabs\+\quad&\cr
                \+single-prime message:  negative $0$-tour\cr
                \smallskip
                \+     &$0 \headmark i'  \Rightarrow  {}\headmark 0\, i$\cr
                \medskip
        \settabs\+\quad&$1\,    0 \headmark j$&\cr
                \+no message:  positive $0$-tour\cr
                \smallskip
                \+     &$\hfill 0 \headmark 1$&${}\Rightarrow  1' \headmark 0$,  propagate into $0$ and then from $0$ to $1$\cr
                \smallskip
                \+     &$1\,    0 \headmark j$&${}\Rightarrow  1\, j'' \headmark 0$, propagate into $0$ and then from $0$ to $1$\quad($j>1$)\cr
                \medskip
        \settabs\+\quad&$0 \headmark j''\, (i+1)\, (i+2)$&\cr
                \+double-prime message:  pushback\cr
                \smallskip
                \+     &$0 \headmark (i+2)''\, (i+1)    $&${}\Rightarrow  0 \headmark (i+1)\, (i+2)'$, propagate from $i+1$ to $i+2$\cr
                \smallskip
                \+     &$0 \headmark j''\, (i+1)\, (i+2)$&${}\Rightarrow  0 \headmark (i+1)\, j''\, (i+2)$, propagate from $i+1$ to 
$i+2$\quad($j>i+2$)\cr}$$

     In our final, unavoidably obscure reformulation, we replace the 
position numbers with nonnumeric base-symbol variables and the 
overarrows that are actually present.  For base-symbol variables whose
overarrows are irrelevant and do not change, however, we omit the
explicit overarrows from the rules.  To avoid explicit reference to
finite-state control, we replace the head marker $\headmark$ with either
$\leftmark$ or $\rightmark$ to indicate whether position $0$ is just to
the left or just to the right.  Except for the start rule (${}\rightmark
a  \Rightarrow  a \leftmark{}$), each rule again has an implicit
symmetric rule.
$$\vbox{\settabs
\+$a     \rvec b$&${}\leftmark c''\rvec d e$&${}\Rightarrow b   \leftmark  \rvec d c'' d$, propagating carry or borrow from $c$ to $b$\cr
\+$\hfill b$     &${}\leftmark c'$          &${}\Rightarrow {}  \rightmark b c$\cr
\medskip
\+$\hfill\rvec b$&${}\leftmark c$           &${}\Rightarrow c'  \rightmark \lvec b$, propagating into $b$, and then from $b$ to $c$\cr
\smallskip
\+$a     \lvec b$&${}\leftmark c$           &${}\Rightarrow ac''\rightmark \lvec b$, propagating into $b$, and then from $b$ to $a$\cr
\medskip
\+$\hfill b$     &${}\leftmark c''\lvec d$  &${}\Rightarrow b   \leftmark  \rvec d c'$, propagating from $d$ to $c$\cr
\smallskip
\+$\hfill b$     &${}\leftmark c''\rvec d e$&${}\Rightarrow b   \leftmark  \rvec d c'' e$, propagating from $d$ to $e$\cr}$$
The start rule closely resembles (the mirror image of) the first of 
the five more general rules.  If we initially provide a singly primed 
``endmarker'' to the left of the head, then the separate start rule 
actually does become redundant; the result, at least if we use radix
$r=4$, is the simulation previewed in Section 2.

\sectionskip\secbegin 8.  Space Analysis//
     The space used for the first $n$ steps of the {\emph most\/} 
space-efficient simulation of $k$ counters is within an additive 
constant of $k \log_2 n$ bits, in the worst case.  For $k$ large, we 
will see now that a straightforward implementation of our real-time,
oblivious simulation requires only about $2.5$ times this much space.

     Regardless of the particular (large enough) radix used, the number 
of distinct {\emph positions\/} involved by step $n$ in the permutation 
process is within an additive constant of $\log_3 n$.  To see this, note 
that the process first reaches position $i+1$ at the end of 
$\tour(i,-)$, and that the number of steps in a negative $i$-tour is 
exactly $(5/4)3^i - (1/2)i - (1/4)$.  The latter, along with the fact 
that the number of steps in a {\emph positive} $i$-tour is exactly 
$(5/4)3^i + (1/2)i - (1/4)$, can be proved by straightforward 
simultaneous induction.

     To minimize the space used for each position, we should choose the 
smallest radix that works.  The analysis below shows that $4$ works.  
For each additional counter, therefore, the space needed for each 
involved position is at most $4=\lceil\log_2 (7 \cdot 2)\rceil$ bits.  
(Each of the seven signed digits has two versions, one underlined and
one not underlined.)  The additional, counter-independent space needed
for each position is at most $3=\lceil\log_2 (3 \cdot 2)\rceil$ bits. 
(The message can be absent, a single prime, or a double prime; and the 
overarrow can point to the left or to the right.)  All together, 
therefore, the space used through step $n$ can be bounded by 
$(3+4k)\log_3 n \approx (1.89 + 2.52 k)\log_2 n$ bits.

     It remains only to show that no overflow (past $3$) or underflow 
(past $\hbox{-3}$) will occur if we use $4$ as the radix in our simulation.  
Until an overflow or underflow does occur, each $(i+1)$-opportunity (and 
also the implicit initialization) will leave each signed digit in 
position $i$ in the range from $\hbox{-2}$ to $2$.  Therefore, it will suffice 
to show that, while there might be as many as four $i$-opportunities 
without an intervening $(i+1)$-opportunity, at most one of these can 
actually result in a carry (or, symmetrically, in a borrow).

     \proclaim Lemma.  For each $i \ge 1$, at most one $i$-opportunity 
in four can result in a carry.  For each $i \ge 0$, therefore, an 
$(i+1)$-opportunity intervenes between every pair of increments to the 
signed digit in position $i$. (Similarly for borrows and decrements, by 
symmetry.)

     \pfbegin Proof.  For each $i \ge 1$, the second assertion follows 
from the first by the third part of the corollary to Observation 5.  For 
$i=0$, the second assertion is an immediate consequence of the second 
part of the same corollary.

     The proof of the first assertion is by induction on $i \ge 1$, and 
the general induction step is itself an induction on time.  Consider the 
first or next $i$-opportunity that results in a carry.  This carry 
leaves the signed digit $\hbox{-1}=3-4$ in position $i-1$.  By the (second) 
assertion for $i-1$, this can increase to at most $0$ by the next 
$i$-opportunity, to at most $1$ by the third $i$-opportunity, and to at 
most $2$ by the fourth $i$-opportunity, none of which requires a
carry.\qed

\sectionskip\secbegin 9.  Further Optimization//
     Our overriding objective so far has been to keep the simulation 
simple. At the expense of some clarity, however, we can make the 
simulation even more efficient.

     There is one easy way to save space in the simulation as presented 
above.  Positions of the separate representations and positions that are 
adjacent in the current permutation need not be encoded separately.  By 
suitable encoding, therefore, the space used can be kept arbitrarily 
close to the unrounded limit $(\log_2 6 + k \log_2 14) \log_3 n \approx 
(1.63 + 2.40 k) \log_2 n$.

     A more subtle observation leads to saving even more space. Because 
each radix-$4$ signed digit is bounded by $3$ in absolute value, the 
number of {\emph significant\/} signed digits in each counter's 
representation stays within an additive constant of the {\emph base-$4$} 
logarithm of the counter's contents.  With care, therefore, we might 
hope to limit the number of positions involved in our simulation to the 
base-$4$ logarithm of the largest counter contents so far.  Even in the 
worst case that the largest counter contents after the first $n$ steps 
is $n$, this would reduce space usage by a factor of $\log_4 n / \log_3 
n \approx .79$.

     One way to take advantage of this potential is to insert some extra 
pairs of negative $i$-tours right before the positive $i$-tour that 
first transports and involves position $i+1$.  (The first half of each 
such pair permutes from $i \ldots 210 \headmark (i+1)(i+2) \ldots$ back 
to the original configuration ${}\headmark 012 \ldots i(i+1)(i+2) 
\ldots\,$, and the second half permutes up to $i \ldots 210 \headmark 
{(i+1)}{(i+2) \ldots}$ again.)  To do this, we need only decide at the time 
we would normally first involve a new position $i+1$ (with a positive 
$0$-tour) whether to start a negative $i$-tour instead (with a {\emph 
negative\/} $0$-tour).  We will want to involve position $i+1$ if and only 
if a significant signed digit is already within a few positions of
position $i+1$.

     For this, we need a second version of each uninvolved position, to 
indicate whether the position is ``ripe'' for involvement, and we need 
appropriate opportunities to mark uninvolved positions ripe.  If $i+1$ 
is the first uninvolved position, then such an opportunity arises each 
time we reach the configuration $(i-3) \ldots 210 \headmark 
(i-2)(i-1)i(i+1) \ldots\,$, say.  It follows from the easy-to-check 
inductive observation below that directional information will suffice to 
identify this situation unambiguously.  If significance has already 
reached position $i-2$, say, then position $i+1$ can be marked finally 
as ripe for involvement, and it will become involved in time to receive 
the first carry from position $i$.

     \proclaim Observation 6.  At any time in the permutation process, 
if $\rvec a b$ occurs anywhere to the right of the head (or, 
symmetrically, if $b \lvec a$ occurs anywhere to the left of the head), 
with a prime or double-prime message attached to neither $a$ nor $b$, 
and if the position number of $a$ is $i$, then the position number of 
$b$ is $i+1$.

     At the expense of obliviousness, this yields a real-time 
multicounter simulation that uses space only logarithmic in the {\emph 
maximum\/} counter contents.  It reduces the worst-case space for a 
real-time simulation of $k$ counters to about $(\log_2 6 + k \log_2 15) 
\log_4 n \approx (1.29 + 1.95 k) \log_2 n$.

     Although, with a slightly different designation of 
$i$-opportunities, we could reduce the radix for our simulation's radix 
notation down to $3$, it turns out to be more space-efficient to use a 
{\emph larger\/} radix.  At the mere expense of additional control states,
this will reduce the number of bits used for underlines, messages, and 
overarrows.  Repeating the analysis sketched above, but now for an 
arbitrary radix $r$, we get a space bound of
$${(\log_2 6 + k \log_2(2(2r-1)+1)) \log_r} n
  \le (k + (\log_2 6 + 2k)/\log_2 r) \log_2 n.$$
For each $\epsilon > 0$, therefore, we can use a radix $r$ so large 
that $(k+\epsilon) \log_2 n$ bits will suffice for every $k$.

     Note that the analyses above do give improved results even for 
oblivious simulation.  Since the counter with the largest contents 
determines head motion, the simulator will be oblivious if it simulates 
one extra, dominant counter of its own, incrementing it at every step.  
This yields a space bound of $(k+1+\epsilon) \log_2 n$ bits for 
oblivious real-time simulation of $k$~counters.

\sectionskip\npoint
\noindent{\tit Acknowledgments.}  We thank Laura Sanchis for
criticizing an early draft of the paper, and Janie Irwin for
providing references [1--2].  P\'eter G\'acs first suggested that the
space used by a nonoblivious version of the simulation could be kept
proportional to the logarithm of the largest counter contents.  The
result of the first 2{,}980{,}000 transitions was obtained by an
implementation of the simulation in PC Scheme, and later checked
independently by John Tromp's implementation in C.\par\tpoint\rm

\sectionskip\secbegin References//
\tpoint\parskip\dd plus 1pt

\def\amm{ The American Mathematical Monthly }
\def\jacm{ Journal of the Association for Computing Machinery }

\def\mst{ Math\-e\-mat\-ical Systems Theory }
\def\jcss{ Journal of Computer and System Sciences }

\def\sicomp{ SIAM Journal on Computing }

\def\iretec{ IRE Trans\-ac\-tions on Electronic Computers }

     1.  D. E. Atkins, {\tit Introduction to the role of redundancy in 
computer arithmetic}, Computer {\bf 8}, 6 (June 1975), 74--77.

     2.  A. Avizienis, {\tit Signed-digit number representations for 
fast parallel arithmetic}, \iretec {\bf EC-10}, 3 (September 1961), 
389--400.

     3.  P. Cull and E. F. Ecklund, Jr., {\tit Towers of Hanoi and
analysis of algorithms}, \amm {\bf 92}, 6 (June--July 1985), 407--420.

     4.  M. J. Fischer and A. L. Rosenberg, {\tit Real-time solutions of 
the origin-crossing problem}, \mst {\bf 2}, 3 (September 1968), 
257--263.

     5.  P. C. Fischer, A. R. Meyer, and A. L. Rosenberg, {\tit Counter 
machines and counter languages}, \mst {\bf 2}, 3 (September 1968), 
265--283.

     6.  P. C. Fischer, A. R. Meyer, and A. L. Rosenberg, {\tit 
Time-restricted sequence generation}, \jcss {\bf 4}, 1 (February 1970), 
50--73.

     7.  M. F\"urer, {\tit Data structures for distributed counting},
\jcss {\bf 28}, 2 (April 1984), 231--243.

     8.  F. C. Hennie and R. E. Stearns, {\tit Two-tape simulation of 
multitape Turing machines}, \jacm {\bf 13}, 4 (October 1966), 533--546.
 
     9.  N. Pippenger and M. J. Fischer, {\tit Relations among 
complexity measures}, \jacm {\bf 26}, 2 (April 1979), 361--381.

     10.  P. M. B. Vit\'anyi, {\tit An optimal simulation of counter 
machines}, \sicomp {\bf 14}, 1 (February 1985), 1--33.

     11.  P. M. B. Vit\'anyi, {\tit An optimal simulation of counter 
machines:  the ACM case}, \sicomp {\bf 14}, 1 (February 1985), 34--40.

\tpoint\parskip 1pt plus 1pt

\bye